\date{}
\documentclass[aps,pra,twocolumn]{revtex4}
\usepackage{float,graphicx,amsmath,amssymb,amsfonts,latexsym,color,dcolumn}
\usepackage{ORI_Group_style}
\newcommand{\PoDF}{\mathbf{r}} 
\newcommand{\PoDFc}{r} 
\newcommand{\MPo}{m} 
\newcommand{\FPo}{\omega_0} 

\newcommand{\PhDF}{\mathbf{R}} 
\newcommand{\PhDFc}{R} 
\newcommand{\MPh}{M} 
 
\newcommand{\CAPo}{q} 

\newcommand{\DampCRR}{\gamma} 

\newcommand{\KerEM}{\mathcal{D}} 

\newcommand{\PropEM}{\mathfrak{D}_{\mu\nu}} 
\newcommand{\PropScalar}{\mathfrak{G}} 

\newcommand{\DampKRRNS}{\Gamma} 

\begin{document}

\title{Radiation Reaction of a Jiggling Dipole in a Quantum Electromagnetic Field}

\author{Adri\'an E. Rubio L\'opez$^{1,2}$\footnote{Adrian.Rubio-Lopez@uibk.ac.at} and Oriol Romero-Isart$^{1,2}$}

\affiliation{$^1$Institute for Quantum Optics and Quantum Information of the Austrian Academy of Sciences, Technikerstraße 21a, Innsbruck 6020, Austria\\$^2$Institute for Theoretical Physics, University of Innsbruck, A-6020 Innsbruck, Austria}


\begin{abstract}
We show how to derive a consistent quantum theory of radiation reaction of a non-relativistic point-dipole quantum oscillator by including the dynamical fluctuations of the position of the dipole. The proposed non-linear theory displays neither runaway solutions nor acausal behaviour without requiring additional assumptions. Furthermore, we show that quantum (zero-point) fluctuations of the electromagnetic field are necessary to fulfil the second law of thermodynamics.

\end{abstract}

\maketitle

A long-standing fundamental problem in electrodynamics is the appearance of runaway and acausal solutions in the dynamics of a moving point charge interacting with its radiated electromagnetic field~\cite{Milonni,Jackson,Oconnell12}.  The so-called radiation reaction problem can be illustrated by the theory of a non-relativistic point-dipole quantum oscillator interacting with the electromagnetic field, whose dynamics is described by the Hamiltonian 
\begin{equation}  \label{eq:H1}
\hat H' = \frac{1}{2 \MPo'} \left[  \hat {\bf p} - q \hat {\bf A} ( \PhDF) \right]^2 + \frac{1 }{2} \kappa   \hat \PoDF  ^2 + \hat H_\text{EM}.
\end{equation}
Here, $\MPo'$ ($\CAPo$) is the bare mass (charge) of the electron, $\CAPo \, \hat \PoDF$ the dipole moment operator, $\kappa$ the observed spring constant of the point-dipole quantum oscillator,  $\hat {\bf A} ( \PhDF) $ the potential vector operator evaluated at the position of the dipole $\PhDF$ (assumed to be fixed), and $\hat H_\text{EM}$ the Hamiltonian describing the dynamics of the free electromagnetic field. From this Hamiltonian, one can readily show~\cite{Milonni,Oconnell12} that the dynamics of the dipole moment degrees of freedom $\PoDF(t) = \langle \hat \PoDF(t) \rangle $ is given by the Abraham-Lorentz~\cite{Abraham02,Lorentz16} equation
\begin{equation}
m \ddot{\PoDF}(t)+ \kappa \PoDF (t) - m \DampCRR\dddot{\PoDF}(t)=0 \, ,
\label{AbrahamLorentz}
\end{equation}
where $m$ is the observed mass of the electron, and $\DampCRR=2\CAPo^{2}/(3\MPo)$   the radiation reaction damping constant. Hereafter natural units are used. The Abraham-Lorentz equation is local in time (Markovian) and in the Fourier space  reads $(\omega^2 - \FPo^2 + \text{i} \DampCRR \omega^3) \rr(\w)=0$, where $\FPo^2 \equiv \kappa/m$ and $\PoDF (t) = (2 \pi)^{-1/2} \int d \omega e^{-\text{i} \omega t }\PoDF(\omega)$.  The radiation reaction problem, that is, the existence of runaways and pre-accelerations, is apparent in the existence of a purely imaginary root of the characteristic polynomial $\omega^2 - \FPo^2 + \text{i} \DampCRR \omega^3$ with positive imaginary part. Historically, the radiation reaction problem is circumvented by either: (i) considering the charge to be extended over a sphere of radius larger than $\gamma$~\cite{Milonni}, (ii) assuming the weak-coupling regime such that one approximates the pathological term $m \dddot{\PoDF}(t) \approx - \kappa \dot{\PoDF}(t)$~\cite{Milonni,BreuerPett}, or (iii) imposing an ultraviolet frequency cut-off in the spectrum of the free electromagnetic field~\cite{FordOconnell91,Oconnell12}.

In this article, we show that by simply promoting the position of the dipole $ \PhDF$ to a {\em dynamical quantum degree of freedom} $\hat \PhDF$, one obtains, without further requirements,  a consistent quantum theory that does not display the radiation reaction problem.
That is, we  describe the radiation reaction of a dipole using the extended Hamiltonian 
\be
\hat H= \hat H' +  \frac{\hat {\bf P}^2}{ 2 \MPh} + V( \hat \PhDF),
\ee 
where  $\hat {\bf P}$ is the conjugate momentum operator of the dipole position operator $\hat \PhDF$, $M$  the total observed mass of the dipole, and $ V$ an external potential for the dipole position degree of freedom. By replacing $\PhDF \rightarrow \hat \PhDF$ in Eq.~(\ref{eq:H1}), $\hat H$ contains a threefold coupling term  between the electromagnetic field, the dipole moment, and its position. In the electric dipole representation, this coupling term reads $\CAPo \, \hat \PoDF \cdot \hat {\bf E}(\hat \PhDF)$, where $\hat {\bf E}$ is the electric field operator.  We  show how to trace out the electromagnetic field and the center-of-mass degrees of freedom in order to get an effective dynamical equation for  the dipole moment degrees of freedom $\hat{\PoDF}(t)$. 
We assume that the center of mass of the dipole does not move on average, $\langle \hat \PhDF (t) \rangle = 0$, but does fluctuate, $\langle \hat \PhDF^2 (t) \rangle \neq 0$. Our analytical procedure leads to a dynamical equation that amends the Abraham-Lorentz Eq.~(\ref{AbrahamLorentz}) and does not suffer from the radiation reaction problem. The amended dynamical equation leads to Eq.~(\ref{AbrahamLorentz}) in the limit $M\rightarrow \infty$, that is, if the center of mass does no longer fluctuate. Furthermore, we show within the developed theory that the zero-point quantum fluctuations of the electromagnetic field are necessary to fulfil the second law of thermodynamics~\cite{FordOconnell88}. 

In the following, we sketch the derivation of our result using the closed-time-path (CTP) formalism and the influence functional method with path integrals~\cite{CalzettaHu}. The detailed derivation can be found in the supplemental material~\cite{SM}.  
 The starting point of the theory is the action of the closed total system, which in the electric dipole representation is given by
\begin{multline}
S\left[\PhDF,\PoDF,A^{\mu}\right]=S_{\rm CM}[\PhDF]+S'_{\rm Dip}[\PoDF]+S_{\rm EM}[A^{\mu}] \\ +S_{\rm Int}\left[\PhDF,\PoDF,A^{\mu}\right].
\label{STotal}
\end{multline}
We use the electromagnetic four-potential $A^{\mu} = ( \phi , {\bf A})$. The first three terms are the actions describing the free dynamics of the subsystems. The action of the dipole center of mass is 
\begin{equation} \label{SCM}
S_{\rm CM}[\PhDF]=\int^{t_{\rm f}}_{t_{\rm in}}d\lambda\left[\frac{\MPh}{2}~\dot{\PhDF}^{2}(\lambda)-V(\PhDF (\lambda))\right].
\end{equation}
The action of the dipole internal degree of freedom is
\begin{equation}
S'_{\rm Dip}[\PoDF]=\int^{t_{\rm f}}_{t_{\rm in}}d\lambda~ \spare{\frac{\MPo}{2} \dot{\PoDF}^{2}(\lambda)-\frac{\kappa'}{2}\PoDF^{2}(\lambda)},
\label{SDip}
\end{equation}
where $\kappa'$ is the bare spring constant of the dipole~\footnote{As opposed to \eqnref{eq:H1}, in our theory it is more convenient to renormalize the spring constant of the dipole than  the mass of the electron.}. The action of the free electromagnetic field is
\begin{equation}
S_{\rm EM}[A^{\mu}]=
\frac{1}{2}\int d^{4}x\left[\mathbf{E}^{2}(x^\mu) - \mathbf{B}^{2} (x^\mu) \right],
\label{SEM}
\end{equation}
where ${\bf B}$ is the magnetic field. We have used the four-vector coordinate $x^\mu = (x^0,{\bf x})$ and the notation  $\int d^{4}x = \int^{t_{\rm f}}_{t_{\rm in}}d x^0 \int d {\bf x}$. The fourth term in \eqnref{STotal} accounts for the threefold interaction between the moving point dipole and the electromagnetic field and is given by
\begin{equation}
S_{\rm Int}\left[\PhDF,\PoDF,A^{\mu}\right]=\CAPo\int^{t_{\rm f}}_{t_{\rm in}}d\lambda~\PoDF (\lambda) \cdot {\bf E}\left(\PhDFc^{\mu}(\lambda)\right),
\label{SInt}
\end{equation}
where $\PhDFc^{\mu}(\lambda)=\left(\lambda,\PhDF(\lambda)\right)$. The initial  state of the total system is assumed to be the product state $\hat{\rho}(t_{\rm in})=\hat{\rho}_{\rm CM}\otimes\hat{\rho}_{\rm Dip}\otimes\hat{\rho}_{\rm EM}$. 

The goal is to obtain the effective equation of motion for the dipole internal degrees of freedom $ \hat \rr$ under the influence of the electromagnetic field and the fluctuations of its center-of-mass position.  For this purpose, we trace out first the electromagnetic field and subsequently the center-of-mass degree of freedom.  The first step benefits from the fact that the  dependence on the electric field in \eqnref{SInt} is linear. The functional integrals that are required to trace out the electromagnetic field within the CTP formalism can be analytically calculated by assuming the initial  state of the electromagnetic field to be Gaussian (\eg~thermal state at temperature $T=\beta^{-1}$), see~\cite{SM,BehuninHu2009,CTPGauge}.

The second step, namely tracing out the center-of-mass degrees of freedom, is more involved due to its non-linear dependence in \eqnref{SInt}. We develop a non-perturbative approximation
technique based on expanding  the influence functional in powers of $q$.
We  perform the functional integrals over the center of mass in the first term of the series. Then, for the $n$th-order term of the series (from two to infinity), we only retain the contribution coming from the $n$th power of the first  term and partially resum the infinite series to obtain an exponential of an effective action. Within the quantum theory of many-particle system with Green's functions, this technique resembles a first-order approximation in the self-energy of an interacting system, which implements a  non-perturbative approximation for the interacting system~\cite{FetterWalecka}. This procedure allows us to obtain a quadratic influence action for the dipole internal degree of freedom, which gives a CTP action of the form
$S_{\rm CTP}[\PoDF,\PoDF']= {S}_{\rm Dip}[\PoDF]- {S}_{\rm Dip}[\PoDF']+S_{\rm RR}[\PoDF,\PoDF']$. Here, ${S}_{\rm Dip}$ has the same form as \eqnref{SDip}, but with the observed (renormalized) spring constant $\kappa' \rightarrow \kappa$. The influence action $S_{\rm RR}$ accounts for the dynamics of both the electromagnetic field and the dipole center-of-mass motion. Although $S_{\rm RR}$ is quadratic, its form is different from what one obtains for standard dissipative environments (\eg,~Brownian motion~\cite{Hu1992, BreuerPett}). Our problem contains a threefold non-linear coupling which leads to a complicated influence action that becomes  quadratic after the non-perturbative method. We remark that this scenario and approach is different from perturbative techniques applied to twofold non-linear interactions~\cite{Hu1993}.

The final step is to obtain the equation of motion for the dipole internal degrees of freedom, which is done by minimizing the CTP action, namely $(\delta S_{\rm CTP}/\delta\PoDF)|_{\PoDF'=\PoDF}=0$~\cite{CalzettaHu}. This leads to the main result of this article, namely the amended Abraham-Lorentz equation
\begin{equation}
m \ddot{\PoDF}(t)+ \kappa \PoDF(t)-2
m \DampCRR \int_{t_{\rm in}}^{t} dt'~D(t-t') \PoDF(t')=0,
\label{RREqCorrected}
\end{equation}
where the memory function is given by 
\begin{widetext} 
\be \label{Kernel0}
D(\tau)=
\int_{0}^{\infty}\frac{d\omega}{(2 \pi)^2}\omega^{3}\exp \spare{-\frac{\omega^{2}}{2}\Delta^{2}(\tau)} \pare{ \cos \spare{ \frac{\omega^{2}}{2}G(\tau)}
{\rm Re}  \left[\DampKRRNS\left(\omega,\tau\right)\right] 
+2\sin \spare{\frac{\omega^{2}}{2}G(\tau)}
{\rm Im}\left[\DampKRRNS\left(\omega,\tau\right)\right]}.
\ee
\end{widetext}
We use $\tau=t-t'$. The memory function includes the effect of the coupling to the electromagnetic field via the function
\be 
\DampKRRNS\left(\omega,\tau \right)\equiv\theta(\tau)\sin \pare{\omega \tau }+\frac{\im }{2}\coth\left(\frac{\beta \omega}{2}\right)\cos \pare{\omega\tau },
\label{eq:Gammafunction}
\ee 
where $\theta(\tau)$ is the Heaviside step function. The imaginary part of \eqnref{eq:Gammafunction} accounts for the fluctuations of the electromagnetic field, where $\coth(\beta \omega/2)=1+2 \bar n (\beta \omega)$ separates the  quantum (zero-point) fluctuations from the classical (thermal) fluctuations.
The  effect of the center-of-mass fluctuations is included in the memory function via  the functions $\Delta^2(t-t')\equiv \tr(\hat{\rho}_{\rm CM}[\hat{\PhDFc}_{j}(t)-\hat{\PhDFc}_{j}(t')]^{2})$ and $G(t-t') = \im \theta(t-t') \tr \pares{ \hat{\rho}_{\rm CM} \coms{\hat{\PhDFc}_{j}(t)}{\hat{\PhDFc}_{j}(t')}}$, which are independent of the axis of motion $j$ and only depend on the time difference for dynamics described by isotroptic and quadratic Hamiltonians. Moreover, $G$ is state independent in this case.

The amended Abraham-Lorentz  \eqnref{RREqCorrected}  contains several features originating from  the non-linear threefold coupling in \eqnref{SInt}. On the one hand, the equation is non-Markovian  with a memory function given by \eqnref{Kernel0}. This is due to the participation of the center of mass in the dipole moment dynamics via the threefold nonlinear coupling that generates delays through the energy exchange between the subsystems. Indeed, in the limit $M \rightarrow \infty$, the functions $\Delta,G$ vanish and $D(t-t') \rightarrow  - \partial^3 \delta(t-t')/\partial t'^3 $. Therefore, in the fixed-dipole limit $M \rightarrow \infty$,   \eqnref{RREqCorrected} leads to the Abraham-Lorentz equation  \eqnref{AbrahamLorentz}.  On the other hand, the damping perceived by the dipole moment depends on the state of the electromagnetic field, namely its temperature, which is a new feature in the radiation reaction scenario. Furthermore, the super-ohmic nature of the electromagnetic field is not altered, as shown by the presence of the factor $\omega^{3}$ in the memory function \eqnref{eq:Gammafunction}. However, the function  $\exp \spare{- \omega^{2} \Delta^2(t-t')/2 }$ in \eqnref{eq:Gammafunction} acts as a  cut-off to the frequency integral since $\Delta^2(t-t') \geq 0$ for any $t-t' \ge 0$. This is a natural cut-off provided by the center-of-mass fluctuations that prevents the localization  of the integral in time, similarly to what is achieved by assuming an ultraviolet frequency cut-off in the electromagnetic field~\cite{FordOconnell91,Oconnell12}.

In order to show that the amended Abraham-Lorentz  \eqnref{RREqCorrected} is free from the radiation reaction problem, it is convenient to write \eqnref{RREqCorrected} in  Fourier space as
\be
\spare{ \w^2  - \w_0^2 + \im \w \mu(\w+ \im 0^+) } \rr(\w) =0.
\ee
Here, the spectral distribution function $ \mu(\w+ \im 0^+) $ is defined as the boundary value on the real axis of the function $\mu(z) \equiv  2 \gamma (\im z)^{-1} \int_0^\infty \text{d} (t-t') D(t-t') e^{\im z (t-t')} $, where we have chosen $t_\text{in} \rightarrow - \infty$. In this form, one can use the results of Ford, Lewis, and O'Connell (FLO)~\cite{FordOconnell88} to show that if $\mu(z)$ is a {\em positive real function}, then the following three general physical principles are fulfilled: (i) causality, which requires $\mu(z)$ to be analytical in the upper half plane $\Im \spares{z} >0$, (ii) the second law of thermodynamics, which enforces the real part of the spectral distribution to be positive $\Re \spares{ \mu(\w+ \im 0^+)} \ge 0$ in all the real axis, and (iii) that $\hat \rr$ is  Hermitian, which requires $\mu(\w+ \im 0^+)= [\mu(-\w+ \im 0^+)]^*$. Note that the Abraham-Lorentz equation \eqnref{AbrahamLorentz} does not fulfil the FLO criteria~\cite{FordOconnell91,Oconnell12}, which is another manifestation of the radiation reaction problem. In contrast, we show in the following that the amended Abraham-Lorentz  \eqnref{RREqCorrected} does fulfil the FLO criteria when the initial states of the center of mass and the electromagnetic field have the same temperature.

In particular, we analytically show that the FLO criteria is met for the paradigmatic case of a free dipole ($V( \RR) = 0$) with an initial motional state $\hat \rho_\text{CM}$ given by a thermal state in a harmonic potential of frequency $\w_I$. Such an initial Gaussian state is determined by $\avg{\hat{\PhDFc}_{i}(t_{\rm in})}=\avg{\hat{P}_{i}(t_{\rm in})}= \avg{\acoms{\hat{R}_{i}(t_{\rm in})}{\hat{P}_{i}(t_{\rm in})}}=0$, and the initial thermal fluctuations $\avg{\hat{\PhDFc}^{2}_{i}(t_{\rm in})}= \spares{2 \bar n(\beta \w_I) +1}/(2 \w_M \w_I)$ and $\avg{\hat{P}^{2}_{i}(t_{\rm in})}= \spares{2 \bar n(\beta \w_I) +1}  \w_M \w_I/2$, where $\w_M \equiv M$ is the Compton frequency. We denote the ratio between the two relevant frequencies describing the  center-of-mass dynamics as $\chi\equiv\w_I/\w_M$. With the center of mass of the dipole being in this initial thermal state, its mean position does not evolve within the free dynamics described by the action \eqnref{SCM} with $V( \RR) = 0$, but it does fluctuate. The thermal wave packet spreads, and this dynamics leads to the following particular expressions for $\Delta^2(t-t') =  \spares{2 \bar n(\beta \w_I) +1} (t-t')^2 \omega_I /(2 \omega_M)$ and $G(t-t')=  \theta (t-t') (t-t')/ \omega_M$. By plugging these functions into \eqnref{Kernel0}, which represent the influence that the center-of-mass degrees of freedom exerts on the dynamics of the dipole moment degrees of freedom, one can  obtain an expression for $\mu(z)$. It can be written as
\begin{equation} \label{eq:mufree}
\mu(z) = \frac{\im \mu_0}{ \pi z} \int_0^\infty \text{d}x x^2 \intall \text{d}y \frac{u(x,y)}{y+z/\w_I}.
\end{equation}
Here, we have defined the positive constant  $\mu_0^{-1} \equiv 4 \sqrt{\pi^3 \chi \coth(\beta \w_I/2)} / (\gamma \w_I^2) $ and the function
$u(x,y)=K_+(x,y)+ \bar n (\beta \w_I x) \spare{K_+(x,y) + K_-(x,y)}$ with
\begin{multline}
K_\pm (x,y) = \exp \spare{- \frac{(y \pm x+ \chi x^2/2)^2}{\chi x^2 \coth(\beta \w_I/2)}} \\ 
- \exp \spare{- \frac{(y \mp x- \chi x^2/2)^2}{\chi x^2 \coth(\beta \w_I/2) }}.
\end{multline}
The spectral distribution $\mu(\w + \im 0^+)$ can be calculated from \eqnref{eq:mufree} using the  distribution identity $i /(x+ \im 0^+) = \im \mathcal{P}(1/x) + \pi \delta(x)$.
One can then readily prove that \eqnref{eq:mufree} meets the FLO criteria by showing that (i) the Cauchy-Riemann equations in the upper half plane are fulfilled, (ii)  the integrand of the real part of the spectral distribution is positive for any value of $\chi$ and $\beta$, and (iii) the real (imaginary) part of the spectral distribution is symmetric (antisymmetric) with respect to $\w$. In accordance with the FLO criteria~\cite{FordOconnell91,Oconnell12}, the amended Abraham-Lorentz equation \eqnref{RREqCorrected} is causal, does not contain runaway solutions, and is consistent with the second law of thermodynamics. While this result has been explicitly (and analytically) shown for the paradigmatic case of a free dipole, it is expected to hold for other center-of-mass dynamics (\eg,~assuming $V(\RR)$ is a harmonic potential).  

To conclude the discussion of the results, let us turn to a subtle but intriguing observation. At finite temperatures, one could be tempted to ignore the quantum (zero-point) fluctuations of the electromagnetic field, in particular since the radiation reaction problem is known to appear also in a classical description of a radiating particle~\cite{Jackson}. The effect of the fluctuations of the electromagnetic field is encoded in the term of the memory function \eqnref{Kernel0} which contains the function $\Im [\Gamma(\w,\tau)]= \coth (\beta \w/2) \cos(\w \tau)/2$. This function can be separated into two terms via $\coth (\beta \w/2) = 1 + 2 \bar n (\beta \w) $. The first term accounts for the quantum (zero-point) fluctuations and the second term for the classical (thermal) fluctuations. Ignoring the quantum (zero-point) fluctuations, namely considering a stochastic classical theory for the electromagnetic field, would result in an amended Abraham-Lorentz equation with the same form as \eqnref{RREqCorrected}, but with a memory function \eqnref{Kernel0} that contains a modified $\Gamma$ function given by $\Gamma_c(\w,\tau) = \theta(\tau) \sin(\w \tau) + \im \cos(\w \tau)/(\beta \w)$, as opposed to \eqnref{eq:Gammafunction}. The Fourier transform of this modified memory function  reads as \eqnref{eq:mufree} but with a modified $u(x,y)$ function given by
\begin{multline}
u_c(x,y) =\frac{1}{2}[K_+(x,y)-K_-(x,y)] \\ +  \frac{1}{\beta \w_I x} \spare{K_+(x,y) + K_-(x,y)}.
\end{multline}
One can then readily show that while the FLO criteria (i) and (iii)  are still fulfilled, and  the theory thus maintains causality, there is a broad range of parameters $\chi$ and $\beta$ for which the FLO criterion (ii) is not fulfilled. Therefore, the quantum (zero-point) fluctuations of the electromagnetic field are necessary to have a theory that is consistent with the second law of thermodynamics. In other words, a radiation reaction theory for a dipole that includes its center-of-mass dynamics would be causal but to be physically consistent, the quantum (zero-point) fluctuations of the electromagnetic field have to be accounted for in order to respect the second law of thermodynamics. We remark that this observation is only pertinent for a moving dipole. In the fixed-dipole limit $M \rightarrow \infty$, the term in the memory function \eqnref{Kernel0}, which contains the relevant function $\Im [\Gamma(\w,\tau)]$ distinguishing classical versus quantum fluctuations of the electromagnetic field, vanishes.

In summary, in this article we have provided an amended Abraham-Lorentz theory for a point-dipole quantum oscillator that is physically consistent and it can be derived from non-relativistic quantum electrodynamics without additional assumptions. The crucial point to circumvent the long-standing radiation reaction problem is to account for the center-of-mass degrees of freedom of the point dipole.  In this context, we have shown that the quantum (zero-point) fluctuations of the electromagnetic field are crucial to obtain a physically consistent theory, even at finite temperatures, since otherwise the theory would violate the second law of thermodynamics. 
From a technical point of view, accounting for the center-of-mass dynamics renders the electrodynamical problem of a point-dipole quantum oscillator interacting with the electromagnetic field non-linear. We were able to obtain analytical results by using path integral techniques~\cite{CalzettaHu}, which are proven to be very well suited to this problem.  

Our results open new research directions. While we have here focused on the effective dynamics of the dipole moment degrees of freedom, it should possible to use similar techniques to describe the effective dynamics of the center-of-mass degrees of freedom. In particular, it might be feasible to obtain a consistent theory that describes the equilibration of the center-of-mass of a dipole interacting with a thermal electromagnetic field, a long-standing open question originally discussed by Einstein and Hopf~\cite{Einstein1910, Lach2012}.  Furthermore,  while we have shown how to circumvent the radiation reaction problem of a point dipole by including additional degrees of freedom, one could investigate whether the same can be achieved for a moving free point charge (\eg, an electron) by including spin degrees of freedom.  Last but not least, it would be very exciting to explore whether such fundamental questions can be addressed experimentally with the new generation of experiments trapping atoms and dielectric nanoparticles in high vacuum near photonic nanostructures, where large light-matter couplings can be engineered.  


\pagebreak
\onecolumngrid

\vspace{1cm}

\begin{center}

   \textbf{\large SUPPLEMENTAL MATERIAL}\\[.2cm]
  \textbf{\large Radiation Reaction of a Jiggling Dipole in a Quantum Electromagnetic Field}\\[.3cm]
  Adri\'an E. Rubio L\'opez$^{1,2}$ and Oriol Romero-Isart$^{1,2}$\\[.1cm]
  {\itshape ${}^1$Institute for Quantum Optics and Quantum Information of the Austrian Academy of Sciences, Technikerstraße 21a, Innsbruck 6020, Austria\\
  ${}^2$Institute for Theoretical Physics, University of Innsbruck, A-6020 Innsbruck, Austria\\}
\end{center}

\setcounter{equation}{0}
\setcounter{figure}{0}
\setcounter{table}{0}
\setcounter{page}{1}
\renewcommand{\theequation}{S\arabic{equation}}
\renewcommand{\thefigure}{S\arabic{figure}}
\renewcommand{\bibnumfmt}[1]{[S#1]}
\renewcommand{\citenumfont}[1]{S#1}

\section{Preliminaries on the path integral formalism}

In the calculation detailed hereafter, we will use the following expression for the action of the free electromagnetic field
\begin{equation}
S_{\rm EM}[A^{\mu}]=-\frac{1}{4}\int d^{4}x~F_{\mu\nu}F^{\mu\nu},
\end{equation}
where $F_{\mu\nu}=\partial_{\mu}A_{\nu}-\partial_{\nu}A_{\mu}$ and $\partial_\mu = \partial/\partial x^\mu$. We employ the metric $\eta_{\mu\nu}=(1,-1,-1,-1)$. 

The action $S_{\rm Int}\left[\PhDF,\PoDF,A^{\mu}\right]$ describing the threefold interaction between the moving point dipole and the electromagnetic field, which is given in the article, can be written as 
\be \label{helpSINT}
S_{\rm Int}\left[\PhDF,\PoDF,A^{\mu}\right]=\int d^{4}x\left[\CAPo\int^{t_{\rm f}}_{t_{\rm in}}d\lambda~\PoDFc^{k}(\lambda) \delta\left(x^{\mu}-\PhDFc^{\mu}(\lambda)\right)\right] E_{k}\left(x^{\mu}\right) = \int d^{4}x~J_{\mu} A^{\mu} = S_{\rm Int}\left[J_\mu, A^\mu \right].
\ee 
The last equation is obtained by integrating by parts and defining the current
\begin{equation} \label{eq:current}
J_{\mu} (x)=-\CAPo\int^{t_{\rm f}}_{t_{\rm in}}d\lambda\left(\partial_{0}\eta_{j\mu}+\partial_{j}\eta_{0\mu}\right)\delta\left(x^{\alpha}-\PhDFc^{\alpha}(\lambda)\right) \PoDFc^{j}(\lambda).
\end{equation}

The fundamental object in the closed-time-path (CTP) formalism is the {\em generating functional}~\cite{SMCalzettaHu}, which in the path integral representation is defined as
\begin{multline}\label{GeneratingFunctionalPathIntegral}
Z[\mathbf{K},\mathbf{K}']=\int d\PhDF_{\rm f} d\PhDF_{\rm in}d\PhDF'_{\rm in} d\PoDF_{\rm f} d\PoDF_{\rm in}d\PoDF'_{\rm in} dA^{\mu}_{\rm f} dA^{\mu}_{\rm in}dA'^{\mu}_{\rm in}~\rho(\PhDF_{\rm in},\PhDF'_{\rm in},\PoDF_{\rm in},\PoDF'_{\rm in},A^{\mu}_{\rm in},A'^{\mu}_{\rm in};t_{\rm in})\\
\times \int^{\PhDF_{\rm f}}_{\PhDF_{\rm in}}\mathcal{D}\PhDF\int^{\PhDF_{\rm f}}_{\PhDF'_{\rm in}}\mathcal{D}\PhDF'\int^{\PoDF_{\rm f}}_{\PoDF_{\rm in}}\mathcal{D}\PoDF\int^{\PoDF_{\rm f}}_{\PoDF'_{\rm in}}\mathcal{D}\PoDF'\int^{A^{\mu}_{\rm f}}_{A^{\mu}_{\rm in}}\mathcal{D}A^{\mu}\int^{A^{\mu}_{\rm f}}_{A'^{\mu}_{\rm in}}\mathcal{D}A'^{\mu} \\ \times \exp \spare{ \im \pare{
S\left[\PhDF,\PoDF,A^{\mu}\right]-S\left[\PhDF',\PoDF',A'^{\mu}\right]+\mathbf{K}\ast\PoDF-\mathbf{K}'\ast \PoDF' }}.
\end{multline}
\noindent Here, we used the notation $\mathbf{A}\ast \mathbf{B}\equiv\int^{t_{\rm f}}_{t_{\rm in}}d\lambda  \mathbf{A}(\lambda)\cdot\mathbf{B}(\lambda)$ and $\rho(\PhDF_{\rm in},\PhDF'_{\rm in},\PoDF_{\rm in},\PoDF'_{\rm in},A^{\mu}_{\rm in},A'^{\mu}_{\rm in};t_{\rm in})=\langle\PhDF_{\rm in},\PoDF_{\rm in},A^{\mu}_{\rm in}|\hat{\rho}(t_{\rm in})|\PhDF'_{\rm in},\PoDF'_{\rm in},A'^{\mu}_{\rm in}\rangle$. The initial initial state is assumed to be in a product state, that is  $\rho(\PhDF_{\rm in},\PhDF'_{\rm in},\PoDF_{\rm in},\PoDF'_{\rm in},A^{\mu}_{\rm in},A'^{\mu}_{\rm in};t_{\rm in})=\rho_{\rm CM}(\PhDF_{\rm in},\PhDF'_{\rm in};t_{\rm in})\rho_{\rm Dip}(\PoDF_{\rm in},\PoDF'_{\rm in};t_{\rm in}) \rho_{\rm EM}(A^{\mu}_{\rm in},A'^{\mu}_{\rm in};t_{\rm in})$. Using this fact, one can write the generating function in terms of the so-called {\em influence functional} $\mathcal{F}_{\rm RR}$, which contains the information of the environment (center of mass and electromagnetic field). It is given by
\begin{multline} \label{ZPo}
Z[\mathbf{K},\mathbf{K}']=\int d\PoDF_{\rm f} d\PoDF_{\rm in}d\PoDF'_{\rm in} \rho_{\rm Dip}(\PoDF_{\rm in},\PoDF'_{\rm in};t_{\rm in}) \\ \times \int^{\PoDF_{\rm f}}_{\PoDF_{\rm in}}\mathcal{D}\PoDF\int^{\PoDF_{\rm f}}_{\PoDF'_{\rm in}}\mathcal{D}\PoDF' 
\exp \spare{ \im 
\pare{ S'_{\rm Dip}\left[\PoDF\right]-S'_{\rm Dip}\left[\PoDF'\right]+\mathbf{K}\ast \PoDF-\mathbf{K}'\ast \PoDF'}}
\mathcal{F}_{\rm RR}[\PoDF,\PoDF'].
\end{multline}
The influence functional can be written as $\mathcal{F}_{\rm RR}= \exp \pare{ \im S_{\rm IF}}$, where $S_\text{IF}$ is defined as the {\em influence action}, and reads
\begin{multline} \label{FRR}
\mathcal{F}_{\rm RR}[\PoDF,\PoDF']=\int d\PhDF_{\rm f} d\PhDF_{\rm in}d\PhDF'_{\rm in}~\rho_{\rm CM}(\PhDF_{\rm in},\PhDF'_{\rm in};t_{\rm in})\int^{\PhDF_{\rm f}}_{\PhDF_{\rm in}}\mathcal{D}\PhDF\int^{\PhDF_{\rm f}}_{\PhDF'_{\rm in}}\mathcal{D}\PhDF' 
\exp \spare{ \im \pare{ 
S_{\rm CM}\left[\PhDF\right]-S_{\rm CM}\left[\PhDF'\right]}}\mathcal{F}_{\rm EM}[J_{\mu},J'_{\mu}],
\end{multline}
where 
\begin{multline}\label{FEMJJ}
\mathcal{F}_{\rm EM}\left[J_{\mu},J'_{\mu}\right]=\int dA^{\mu}_{\rm f} dA^{\mu}_{\rm in}dA'^{\mu}_{\rm in}~\rho_{\rm EM}(A^{\mu}_{\rm in},A'^{\mu}_{\rm in};t_{\rm in})
\\ \times
\int^{A^{\mu}_{\rm f}}_{A^{\mu}_{\rm in}}\mathcal{D}A^{\mu}\int^{A^{\mu}_{\rm f}}_{A'^{\mu}_{\rm in}}\mathcal{D}A'^{\mu}  
\exp \spare{ \im \pare{ 
S_{\rm EM}\left[A^{\mu}\right]+S_{\rm Int}\left[J_{\mu},A^{\mu}\right]-S_{\rm EM}\left[A'^{\mu}\right]-S_{\rm Int}\left[J'_{\mu},A'^{\mu}\right]}}
\end{multline}
is the influence functional containing the information of the electromagnetic field.

In \secref{sec:EM} we will first perform the integrals in \eqnref{FEMJJ} (trace out the electromagnetic field).  In \secref{sec:EM} we will then perform the integrals in \eqnref{FRR} (trace out the center of mass), which will lead to  an expression for the influence action $S_\text{IF}[\rr,\rr']$. In this form, the generating function will be given by
\be \label{ZPointegrated} 
Z[\mathbf{K},\mathbf{K}']=\int d\PoDF_{\rm f} d\PoDF_{\rm in}d\PoDF'_{\rm in} \rho_{\rm Dip}(\PoDF_{\rm in},\PoDF'_{\rm in};t_{\rm in})  \times \int^{\PoDF_{\rm f}}_{\PoDF_{\rm in}}\mathcal{D}\PoDF\int^{\PoDF_{\rm f}}_{\PoDF'_{\rm in}}\mathcal{D}\PoDF' 
\exp \spare{ \im S_\text{CTP} [\PoDF,\PoDF']},
\ee
where we define the CTP action
\be 
 S_{\rm CTP}[\PoDF,\PoDF'] = S'_{\rm Dip}\left[\PoDF\right]-S'_{\rm Dip}\left[\PoDF'\right] + S_\text{IF}[\PoDF,\PoDF']+\mathbf{K}\ast \PoDF-\mathbf{K}'\ast \PoDF'.
\ee 
Note that the generating function in \eqnref{ZPointegrated} contains only integrals over the dipole moment degrees of freedom. At this point, one can then calculate the amended Abraham-Lorentz equation, that is,    the effective dynamical equation for $\rr$, via the minimization of the CTP action~\cite{SMCalzettaHu}, namely
\be \label{eq:eom}
\left. \frac{\delta S_{\rm CTP}[\PoDF,\PoDF']}{\delta\PoDF(t)}\right|_{\PoDF=\PoDF', \KK=\KK'=0} = 0.  
\ee

\section{Tracing out the electromagnetic field} \label{sec:EM}

Let us trace out first the electromagnetic field by performing the integrals in the functional \eqnref{FEMJJ}. We remark that the path integrals of the electromagnetic field have to exclude Gauge equivalent paths, something that is done using the Faddeev-Popov procedure~\cite{SMCTPGauge}.  Regardless on the chosen Gauge, the integral is Gaussian and one obtains~\cite{SMBehuninHu2009}
\begin{equation}
\mathcal{F}_{\rm EM}\left[J^{\mu-},J^{\mu+}\right]= \exp \cpare{ \im 
\int d^{4}y\int d^{4}y'~J^{\mu-}(y)\left[2\PropEM^{\rm Ret}(y-y')~J^{\nu+}(y')+\frac{i}{2}\PropEM^{\rm H}(y-y')~J^{\nu-}(y')\right]} = e^{\im S_\text{IEM}},
\label{FEMJJSolved}
\end{equation}
\noindent where $J^{\nu+}=(J^{\nu}+J'^{\nu})/2$, $J^{\nu-}=J^{\nu}-J'^{\nu}$, and $S_\text{IEM}$  is the influence action due to the electromagnetic field. We have also defined $\PropEM^{\rm Ret}$ ($\PropEM^{\rm H}$) as the retarded (Hadamard) propagator of the electromagnetic field in the thermal state of temperature $T$. These propagators are calculated in the chosen Gauge where the functional integration was performed. Following \cite{SMBehuninHu2009}, one can express the propagators in the Feynman Gauge, namely $\PropEM^{\rm Ret}(y-y')= \im \theta(y_{0}-y'_{0}) \tr \spares{ \coms{\hat A_\mu (y)}{\hat A_\nu (y')} \hat \rho_\text{EM}(t_\text{in})}=\eta_{\mu\nu}\PropScalar_{\rm Ret}(y-y')$ and $\PropEM^{\rm H}(y-y')=\tr \spares{ \acoms{\hat A_\mu (y)}{\hat A_\nu (y')} \hat \rho_\text{EM}(t_\text{in})}=\eta_{\mu\nu}\PropScalar_{\rm H}(y-y')$, where
\begin{equation}
\PropScalar_{\rm Ret}(y-y')=- \theta(y_{0}-y'_{0})\int\frac{d\mathbf{p}}{(2\pi)^{3}}~e^{\im\mathbf{p}\cdot\left(\mathbf{y}-\mathbf{y}'\right)} \frac{\sin\left[\omega_{\mathbf{p}}(y_{0}-y'_{0})\right]}{\omega_{\mathbf{p}}},
\label{RetKernelEM}
\end{equation}
and
\begin{equation}
\PropScalar_{\rm H}(y-y')=\int\frac{d\mathbf{p}}{(2\pi)^{3}}~e^{\im\mathbf{p}\cdot\left(\mathbf{y}-\mathbf{y}'\right)}~\coth\left(\frac{\beta \omega_{\mathbf{p}} }{2}\right)\frac{\cos\left[\omega_{\mathbf{p}}(y_{0}-y'_{0})\right]}{\omega_{\mathbf{p}}},
\label{HKernelEM}
\end{equation}
\noindent with $\omega_{\mathbf{p}}=|\mathbf{p}|$ and $\beta=1/T$. Note that $\mathcal{F}_{\rm EM}$ is a functional of $\PhDF,\PhDF',\PoDF,\PoDF'$ through its dependence on the four-currents $J^{\mu}$ given in \eqnref{eq:current}. 

Then, by making use of \eqnref{eq:current}, \eqnref{RetKernelEM} and \eqnref{HKernelEM}, the exponent of \eqnref{FEMJJSolved} can be written as 
\begin{multline}
    S_\text{IEM}\left[\PoDF,\PoDF',\PhDF,\PhDF'\right]= 
\int_{t_{\rm in}}^{t_{\rm f}} d\lambda d\lambda' \Big\{\PoDFc^{j}(\lambda)\PoDFc^{k}(\lambda') 
\spare{\KerEM_{jk}^{\rm Ret} \pare{\lambda-\lambda',\PhDF(\lambda)-\PhDF(\lambda')} +\frac{i}{2}\KerEM_{jk}^{\rm H} (\lambda-\lambda',\PhDF(\lambda)-\PhDF(\lambda')) }\\
+\PoDFc^{j}(\lambda) \PoDFc'^{k}(\lambda') \spare{\KerEM_{jk}^{\rm Ret} (\lambda-\lambda',\PhDF(\lambda)-\PhDF'(\lambda')) -\frac{i}{2}\KerEM_{jk}^{\rm H} (\lambda-\lambda',\PhDF(\lambda)-\PhDF'(\lambda')) } \\
-\PoDFc'^{j} (\lambda) \PoDFc^{k}(\lambda')\spare{\KerEM_{jk}^{\rm Ret} (\lambda-\lambda',\PhDF'(\lambda)-\PhDF(\lambda'))+\frac{i}{2}\KerEM_{jk}^{\rm H} (\lambda-\lambda',\PhDF'(\lambda)-\PhDF(\lambda'))}\\
-\PoDFc'^{j}(\lambda) \PoDFc'^{k}(\lambda')\spare{\KerEM_{jk}^{\rm Ret}(\lambda-\lambda',\PhDF'(\lambda)-\PhDF'(\lambda')) -\frac{i}{2}\KerEM_{jk}^{\rm H}(\lambda-\lambda',\PhDF'(\lambda)-\PhDF'(\lambda'))}\Big\},
\end{multline}
where
\begin{multline}
\KerEM^{\text{Ret}}_{jk}(\lambda-\lambda',\PhDF(\lambda)-\PhDF'(\lambda'))=\frac{\CAPo^{2}}{2}\left[\left(-\partial_{0}\delta_{j}^{\mu}+\partial_{j}\delta_{0}^{\mu}\right)\left(-\partial'_{0}\delta_{k}^{\nu}+\partial'_{k}\delta_{0}^{\nu}\right)\PropEM^{\text{Ret}}(y,y')\right]\Big|_{y^{\alpha}=(\lambda,\PhDF(\lambda)),y'^{\alpha}=(\lambda',\PhDF'(\lambda'))}\\
= -\frac{\CAPo^{2}}{2}\delta_{jk}\delta(\lambda-\lambda')\int\frac{d\mathbf{p}}{(2\pi)^{3}}~e^{\im\mathbf{p}\cdot\left(\PhDF(\lambda)-\PhDF'(\lambda)\right)}
\\+\frac{\CAPo^{2}}{2}\theta(\lambda-\lambda')\int\frac{d\mathbf{p}}{(2\pi)^{3}}\left[\omega_{\mathbf{p}}^{2}\delta_{jk}-p_{j}p_{k}\right]e^{\im\mathbf{p}\cdot\left(\PhDF(\lambda)-\PhDF'(\lambda')\right)}~\frac{\sin\left[\omega_{\mathbf{p}}(\lambda-\lambda')\right]}{\omega_{\mathbf{p}}},
\end{multline}
and
\begin{multline}
\KerEM^{\text{H}}_{jk}(\lambda-\lambda',\PhDF(\lambda)-\PhDF'(\lambda'))\equiv\frac{\CAPo^{2}}{2}\left[\left(-\partial_{0}\delta_{j}^{\mu}+\partial_{j}\delta_{0}^{\mu}\right)\left(-\partial'_{0}\delta_{k}^{\nu}+\partial'_{k}\delta_{0}^{\nu}\right)\PropEM^{\text{H}}(y,y')\right]\Big|_{y^{\alpha}=(\lambda,\PhDF(\lambda)),y'^{\alpha}=(\lambda',\PhDF'(\lambda'))}\\
= \frac{\CAPo^{2}}{2}\int\frac{d\mathbf{p}}{(2\pi)^{3}}\left[\omega_{\mathbf{p}}^{2}~\delta_{jk}-p_{j}p_{k}\right]e^{\im\mathbf{p}\cdot\left(\PhDF(\lambda)-\PhDF'(\lambda')\right)}\coth\left(\frac{\beta \omega_{\mathbf{p}}}{2}\right)~\frac{\cos\left[\omega_{\mathbf{p}}(\lambda-\lambda')\right]}{\omega_{\mathbf{p}}}.
\end{multline}

The section is concluded by rewriting $S_\text{IEM}$ such that its dependence on the center of mass variables is explicit, namely
\begin{multline} \label{helpSIEM}
S_\text{IEM}\left[\PoDF,\PoDF',\PhDF,\PhDF'\right]=
S^{0}_\text{IEM}[\rr,\rr']
+\int_{t_{\rm in}}^{t_{\rm f}} d\lambda d\lambda'~\frac{\CAPo^{2}}{2}\int\frac{d\mathbf{p}}{(2\pi)^{3}}\frac{\left(\omega_{\mathbf{p}}^{2}~\delta_{jk}-p_{j}p_{k}\right)}{\omega_{\mathbf{p}}} \times \\
\times\Bigg[\PoDFc^{j}(\lambda) \PoDFc^{k}(\lambda') e^{\im\mathbf{p}\cdot \spare{\PhDF(\lambda)-\PhDF(\lambda')}}\DampKRRNS\left(\omega_{\mathbf{p}},\lambda-\lambda'\right)
+\PoDFc^{j}(\lambda) \PoDFc'^{k}(\lambda') e^{\im\mathbf{p}\cdot \spare{\PhDF(\lambda)-\PhDF'(\lambda')}}\DampKRRNS^{\ast}\left(\omega_{\mathbf{p}},\lambda-\lambda'\right)
\\
-\PoDFc'^{j}(\lambda) \PoDFc^{k}(\lambda')  e^{\im\mathbf{p}\cdot \spare{\PhDF'(\lambda)-\PhDF(\lambda')}}\DampKRRNS\left(\omega_{\mathbf{p}},\lambda-\lambda'\right)
-\PoDFc'^{j}(\lambda) \PoDFc'^{k}(\lambda') e^{\im\mathbf{p}\cdot \spare{\PhDF'(\lambda)-\PhDF'(\lambda')}}\DampKRRNS^{\ast}\left(\omega_{\mathbf{p}},\lambda-\lambda'\right)
\Bigg],
\end{multline}
where we have defined
\begin{equation} 
\DampKRRNS\left(\omega_{\mathbf{p}},\lambda -\lambda'\right)\equiv\theta(\lambda -\lambda')\sin \spare{\omega_{\mathbf{p}}\pare{\lambda -\lambda'}}+\frac{i}{2}\coth\left(\frac{\beta \omega_{\mathbf{p}}}{2}\right)\cos \spare{\omega_{\mathbf{p}}\pare{\lambda -\lambda'}},
\label{CorrEMFree}
\end{equation}
and
\be \label{S0IEM}
S^{0}_\text{IEM}[\rr,\rr']=-\frac{\CAPo^{2}}{2}\int_{t_{\rm in}}^{t_{\rm f}} d\lambda \int\frac{d\mathbf{p}}{(2\pi)^{3}}\left[\PoDF^{2}(\lambda)-\PoDF'^{2}(\lambda)\right] = -\CAPo^{2}\int_{t_{\rm in}}^{t_{\rm f}} d\lambda \int_0^\infty \frac{d \w_\mathbf{p}}{(2\pi)^{2}} \w^2_\pp \left[\PoDF^{2}(\lambda)-\PoDF'^{2}(\lambda)\right].
\ee
Note that $S^{0}_\text{IEM}$, which depends only on the dipole moment coordinates, is a divergent term that   renormalizes the spring constant of the dipole degree of freedom, see \eqnref{eq:kappa}. Hereafter it is convenient to define $\tilde S_\text{IEM} = S_\text{IEM} - S^{0}_\text{IEM} $.

\section{Tracing out the center-of-mass degrees of freedom: non-perturbative approximation} \label{sec:CM}

This section focuses on the integration of the center-of-mass degrees of freedom in the influence functional $\mathcal{F}_\text{RR}$, see \eqnref{FRR}. Note that this integration cannot be done analytically since $\mathcal{F}_\text{RR}$ depends on $\mathcal{F}_\text{EM} = e^{\im S_\text{IEM}}$ (see \eqnref{FRR}) and $S_\text{IEM}$ contains the coordinates of the center of mass in exponents (see \eqnref{helpSIEM}). This is a consequence of the non-linear coupling with the center of mass in $S_\text{Int}$ (see \eqnref{helpSINT}). To proceed, we   expand $\mathcal{F}_\text{EM}$ in a Taylor series  around zero coupling, namely around $q=0$, that is
\be \label{FIEMexpanded} 
\mathcal{F}_\text{EM} = e^{\im S_\text{IEM}} =e^{\im S^0_\text{IEM}} \sum_{n=0} \frac{(\im q)^n}{n!} \frac{\partial^n \pare{ \tilde S_\text{IEM}}}{\partial q^n}. 
\ee 
Note that the action $S^{0}_\text{IEM}$, see \eqnref{S0IEM}, which does not depend on the center-of-mass coordinates, is factorized from the expansion. By plugging \eqnref{FIEMexpanded} into  \eqnref{FRR}, one obtains a series expansion for $\mathcal{F}_\text{RR}$, namely
\begin{multline}
\mathcal{F}_{\rm RR}[\PoDF,\PoDF']=e^{\im S^0_\text{IEM}[\rr,\rr']} \sum_{n=0}\frac{(\im q)^n}{n!}\int d\PhDF_{\rm f} d\PhDF_{\rm in}d\PhDF'_{\rm in} \rho_{\rm CM}(\PhDF_{\rm in},\PhDF'_{\rm in};t_{\rm in}) \times \\ \times \int^{\PhDF_{\rm f}}_{\PhDF_{\rm in}}\mathcal{D}\PhDF\int^{\PhDF_{\rm f}}_{\PhDF'_{\rm in}}\mathcal{D}\PhDF' 
e^{ \im \pare{ 
S_{\rm CM}\left[\PhDF\right]-S_{\rm CM}\left[\PhDF'\right]}}\frac{\partial^n \pare{ S_\text{IEM}}}{\partial q^n} =e^{\im  S^0_\text{IEM}[\rr,\rr']} \pare{ 1 + \sum_{n=1} \im^n \mathcal{S}^{(n)}_\text{RR} [\PoDF,\PoDF']}.
\end{multline}

One can now evaluate $\mathcal{S}^{(1)}_\text{RR} [\PoDF,\PoDF']$ by using the following relation~\cite{SMCalzettaHu}
\begin{multline} \label{RIntegration}
\int d\mathbf{R}_{\rm f}  d\mathbf{R}_{\rm in}d\mathbf{R}'_{\rm in}~\rho_{\rm CM}(\mathbf{R}_{\rm in},\mathbf{R}'_{\rm in};t_{\rm in})\int^{\mathbf{R}_{\rm f}}_{\mathbf{R}_{\rm in}}\mathcal{D}\mathbf{R}\int^{\mathbf{R}_{\rm f}}_{\mathbf{R}'_{\rm in}}\mathcal{D}\mathbf{R}'e^{\im
\left(S_\text{CM}\left[\mathbf{R}\right]-S_\text{CM}\left[\mathbf{R}'\right]+ \mathbf{K} \ast \mathbf{R} - \mathbf{K}'\ast \mathbf{R}'\right)}\\
=\exp \spare{\im\int_{t_{\rm in}}^{t_{\rm f}}d\lambda d \lambda'\mathbf{K}^-(\lambda)\cdot\left(G(\lambda-\lambda') \mathbf{K}^+(\lambda')+\frac{i}{4}G_\text{H}(\lambda-\lambda') \mathbf{K}^-(\lambda')\right)}.
\end{multline}
It is assumed that the center-of-mass free dynamics is quadratic and isotropic such that  $G(\lambda-\lambda')= \im \theta(\lambda-\lambda') \tr \spares{ \coms{\hat R_i(\lambda)}{\hat R_i (\lambda')} \hat \rho_\text{CM}(t_\text{in})} $ and $G_\text{H}(\lambda-\lambda') =  \tr \spares{ \acoms{\hat R_i(\lambda)}{\hat R_i (\lambda')} \hat \rho_\text{CM}(t_\text{in})} $ for any $i=x,y,z$. Using \eqnref{RIntegration} and recalling \eqnref{helpSIEM}, one can the obtain
\begin{multline}
\mathcal{S}^{(1)}_{\rm RR}\left[\PoDF,\PoDF'\right]
=\frac{2  q^2}{3 \pi}
\int_{t_{\rm in}}^{t_{\rm f}} d\lambda d\lambda'\int_{0}^{\infty}\frac{d\omega_{\mathbf{p}}}{2\pi}
\omega_{\mathbf{p}}^{3}\nonumber
\exp \spare{-\frac{\omega_{\mathbf{p}}^{2}}
{2}\Delta^{2}\left(\lambda-\lambda'\right)}
\times \\ \times 
\Bigg \{ \PoDF(\lambda)\cdot\PoDF(\lambda') \exp \spare{-\im \frac{\omega_{\mathbf{p}}^{2}}{2}\left(G(\lambda-\lambda')+G(\lambda'-\lambda)\right)}\DampKRRNS\left(\omega_{\mathbf{p}},\lambda-\lambda'\right)
\\
+ \PoDF(\lambda)\cdot\PoDF'(\lambda') \exp \spare{\im\frac{\omega_{\mathbf{p}}^{2}}{2}\left(G(\lambda-\lambda')-G(\lambda'-\lambda)\right)}\DampKRRNS^{\ast}\left(\omega_{\mathbf{p}},\lambda-\lambda'\right)
\\
-\PoDF'(\lambda)\cdot\PoDF(\lambda') \exp \spare{-\im \frac{\omega_{\mathbf{p}}^{2}}{2}\left(G(\lambda-\lambda')-G(\lambda'-\lambda)\right)}\DampKRRNS\left(\omega_{\mathbf{p}},\lambda-\lambda'\right)
\\
-\PoDF'(\lambda)\cdot\PoDF'(\lambda') \exp \spare{\im\frac{\omega_{\mathbf{p}}^{2}}{2}\left(G(\lambda-\lambda')+G(\lambda'-\lambda)\right)}\DampKRRNS^{\ast}\left(\omega_{\mathbf{p}},\lambda-\lambda'\right)
\Bigg \},
\end{multline}
where we have defined  $\Delta^2(\lambda-\lambda')=[G_\text{H}(\lambda,\lambda)+G_\text{H}(\lambda',\lambda')-2G_\text{H}(\lambda,\lambda')]/2=\tr \pares{\hat{\rho}_{\rm CM}[\hat{R}(\lambda)-\hat{R}(\lambda')]^{2}} \ge 0$. Note that the functions $\Delta(t-t')$ and $G(t-t')$ appeared in the the final result discussed in the main text of the article.

It is at this point where we perform the following non-perturbative approximation to calculate $\mathcal{F}_\text{RR}$,
\be  
\mathcal{F}_\text{RR}  = e^{\im  S^0_\text{IEM} } \pare{ 1 + \sum_{n=1} \im^n \mathcal{S}^{(n)}_\text{RR} } \approx e^{\im  S^0_\text{IEM} } \sum_{n=0} \frac{1}{n!} \spare{ \im \mathcal{S}^{(1)}_{\rm RR}  }^n = \exp \spare{ \im \pare{S^0_\text{IEM}  + \mathcal{S}^{(1)}_{\rm RR}}}.
\ee
That is, we approximate $\mathcal{S}^{(n)}_\text{RR} \approx \spare{\mathcal{S}^{(1)}_{\rm RR}}^n/n!$.
Within this approximation, the CTP action is given by
\begin{multline} 
S_\text{CTP}[\rr,\rr'] \approx S'_{\rm Dip}\left[\PoDF\right]-S'_{\rm Dip}\left[\PoDF'\right]+S^0_\text{IEM}[\rr,\rr']  + \mathcal{S}^{(1)}_{\rm RR}[\rr,\rr']+\mathbf{K}\ast \PoDF-\mathbf{K}'\ast \PoDF' \\
= S_{\rm Dip}\left[\PoDF\right]-S_{\rm Dip}\left[\PoDF'\right]  + \mathcal{S}^{(1)}_{\rm RR}[\rr,\rr']+\mathbf{K}\ast \PoDF-\mathbf{K}'\ast \PoDF',
\end{multline}
where as defined in the article $S_{\rm Dip}$ is obtained from  $S_{\rm Dip}'$ by renormalizing the spring constant by
\be \label{eq:kappa}
\kappa = \kappa' +  q^2 \int_0^{\infty}\frac{d \w_\pp}{(2 \pi)^2} \w_\pp^2.
\ee 

We remark that although the form of the influence action is quadratic in the dipole variables, it is formally different from the one obtained in the quantum Brownian motion theory~\cite{SMHu1992}. This can be clearly observed by writing $S^{(1)}_\text{RR}$ in terms of the sum and difference variables $\PoDF^{+}=(\PoDF+\PoDF')/2$, $\PoDF^{-}=\PoDF-\PoDF'$. All the therms containing products of $\PoDF^{+}$ and $\PoDF^{-}$ are present, while in the Brownian motion theory the products $\PoDF^{+}\cdot\PoDF^{+}$ do not appear~\cite{SMHu1992}.

The amended Abraham-Lorentz equation for the dipole moment is then obtained using \eqnref{eq:eom}, which leads to
\begin{multline} \label{eq:ALresult}
m \ddot{\PoDF}(t)+ \kappa \PoDF(t)-2 m \DampCRR \int_{t_{\rm in}}^{t} d\lambda
\int_{0}^{\infty}\frac{d\omega_{\mathbf{p}}}{(2 \pi)^2}\omega_{\mathbf{p}}^{3}\exp \spare{-\frac{\omega_{\mathbf{p}}^{2}}{2}\Delta^{2}(t-\lambda)} \times \\ 
\times \Bigg \{ \cos \spare{ \frac{\omega_{\mathbf{p}}^{2}}{2}G(t-\lambda)}
{\rm Re}  \left[\DampKRRNS\left(\omega_{\mathbf{p}},t-\lambda\right)\right] 
+2\sin \spare{\frac{\omega_{\mathbf{p}}^{2}}{2}G(t-\lambda)}
{\rm Im}\left[\DampKRRNS\left(\omega_{\mathbf{p}},t-\lambda\right)\right]\Bigg \} \PoDF(\lambda)=0.
\end{multline}
Using the expression of $\Gamma(\w,t-t')$ obtained in \eqnref{CorrEMFree} leads to result shown in the main article.

\end{document}